\documentclass{rmf-d}
\usepackage{nopageno,rmfbib,multicol,times,epsf,amsmath,amssymb,cite}
\usepackage[latin1]{inputenc}
\usepackage[]{caption2}
\usepackage{graphics}
\usepackage[demo]{graphicx}
\usepackage{eucal}
\usepackage{amsmath,amssymb}
\usepackage{hyperref}
\usepackage{epstopdf}
\usepackage{slashed}
\usepackage{ulem}
\def\rmfcornisa{
}

\newenvironment{Figure}
  {\par\medskip\noindent\minipage{\linewidth}}
  {\endminipage\par\medskip}%
\def\rmfcintilla{
}
\clearpage \rmfcaptionstyle 
\begin{document}
\title{   The $\tau\to\nu_\tau\pi e^+e^-$ decay revisited
\vspace{-6pt}}
\author{ Adolfo Guevara}
\address{ Departamento de F\'isica At\'omica, Molecular y Nuclear
and Instituto Carlos I de F\'isica Te\'orica y Computacional
Universidad de Granada, E-18071 Granada, Spain. }
\author{ }
\address{Departament de F\'isica Te\`orica, IFIC, Universitat de Val\`encia - CSIC,
Apt. Correus 22085, E-46071 Val\`encia, Spain }
\author{ }
\address{ }
\author{ }
\address{ }
\author{ }
\address{ }
\author{ }
\address{ }
\maketitle
\begin{abstract}
\vspace{1em} {\bf Abstract.} The recent results from Belle of the $\tau^-\to\nu_\tau\pi^-e^+e^-$ analysis has incentivated us to make a reanalysis of a former work where the Structure Dependent parts are obtained using Resonance Chiral Theory. Here we rely on the same theory accounting for effects that explicitly break such symmetry. A motivation is the involved $W\pi\gamma^\star$ vertex, utterly relevant in computing $\tau^- \to \nu_{\tau}\pi^-$ radiative corrections.  
\end{abstract}
\keys{  Resonance Chiral Theory, Chiral Lagrangians, hadronic tau decays, Belle physics  \vspace{-4pt}}
\pacs{   \bf{\textit{12.39.Fe, 12.38-t, 11.15.Pg}}    \vspace{-4pt}}
\begin{multicols}{2}

\section{Introduction}

(This work is based on the research of ref. \cite{Guevara:2021tpy}.) The increasing development in the search for Beyond Standard Model Physics in flavor factories demands a better control of backgrounds, specially in processes allowed in the Standard Model with large suppressions. It has been shown that, to the precision achieved in such experiments, important background may arise from radiative corrections \cite{Kampf:2018wau,Husek:2017vmo,Guevara:2016trs} and hadronic contamination \cite{Guevara:2015pza}. In particular, $\tau$ decays provide a suitable scenario to search for BSM phenomena since they involve low hadronic contamination compared to purely hadronic transitions and since at experiments like Belle, a very large amount of them are created. Therefore, such decays are suitable to look for BSM effects, such as Lepton Flavor- Lepton Number and Lepton Universality-Violation, in semileptonic $\tau$ decays. Some of the aforementioned effects could induce decays such as $\tau^-\to\nu_\tau\pi^+\mu^-\mu^-$ or $\tau^-\to\mu^-\gamma$ that could be explored in Belle-II. In this work, we focus on an important background of such processes, namely, the $\tau^-\to\nu_\tau \pi^-\ell\overline{\ell}$. Furthermore, the relevance of such decays relies on the effective vertex $W\gamma^* P$, which is of utmost importance in the evaluation of radiative corrections to the $\tau\to\nu_\tau P$ \cite{Guo:2010dv} decays as well as in determining high energy behavior of the Transition Form Factor (TFF) of the $\pi$ meson, which is needed to describe the most important parts of the Hadronic Vacuum Polarization contribution to the anomalous magnetic moment of the $\mu$.

The recent measurement of the $\tau^-\to\nu_\tau \pi^-e^+e^-$ decays in Belle \cite{Jin:2019goo} has confirmed our previous determination of the Branching Fraction of such process \cite{Guevara:2013wwa}, where we relied on Resonance Chiral Theory (R$\chi$T) \cite{Ecker:1988te,Ecker:1989yg} (which is $\chi$PT \cite{Weinberg:1978kz,Gasser:1983yg,Gasser:1984gg} expanded to include resonances as active degrees of freedom) , where the lack of a pseudoscalar-resonance exchange and some $\mathcal{O}(p^6)$ contributions provide an incomplete axial form factor in the sense of $VAP$ Green's function analysis \cite{Guevara:2021tpy,Cirigliano:2004ue,Cirigliano:2006hb}. This motivated us to revisit the analysis of such decays taking into account the missing parts in the form factors by including the pseudoscalar resonance exchange. These resonances, however, will necessarily couple to a pseudo Goldstone boson and such coupling involves the mass of the latter. Therefore, these resonances will contribute through terms that break explicitly the Chiral Symmetry\footnote{Terms in the effective Lagrangian that depend on the masses of the quarks (or equivalently on the masses of the pseudo Goldstone bosons) will break chiral symmetry explicitly. Since the chiral symmetry is spontanously broken $U(3)_L\otimes U(3)_R\to U(3)_V$, considering quark-mass terms means that the $SU(3)_V$ subgroup gets explicitly broken.}. Thus, for the sake of consistency 
(to see how the $VVP$ Green's function analysis \cite{RuizFemenia:2003hm,Kampf:2011ty} relates to the  vector form factor see ref.\cite{Guevara:2018rhj}), all the leading-order contributions ($\mathcal{O}(m_\pi^2)$) to the remaining $U(3)_V$ breaking must be considered. Therefore, our analysis will involve the missing $\mathcal{O}(p^6)$ terms in ref. \cite{Guevara:2013wwa} as well as all leading mass contributions to the $\tau^-\to\nu_\tau \pi^-\ell\overline{\ell}$ decays. (Although we also analyzed the $\tau^-\to\nu_\tau K^-\ell\overline{\ell}$ in ref. \cite{Guevara:2021tpy}, here we focus only in the $\pi$ decay-channel.)

\begin{Figure}
 \centering\includegraphics[width=\linewidth]{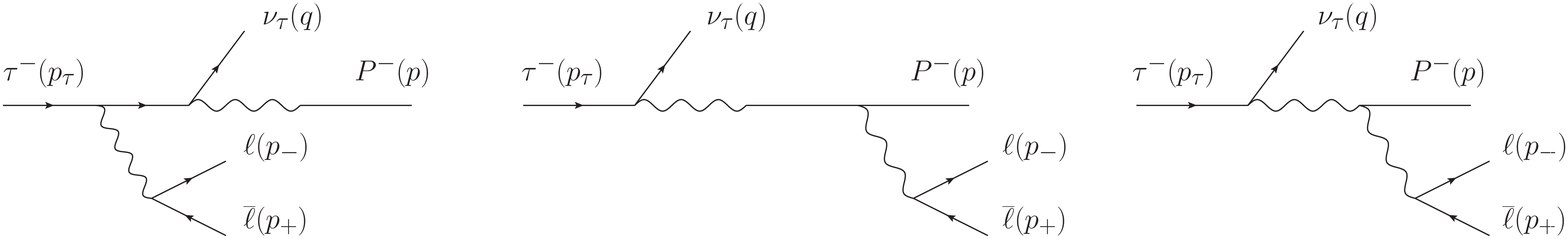}\label{fig:SI}\\
 \footnotesize{Figure 1. Feynman diagrams of the SI contributions.}
\end{Figure}

\section{Amplitudes}

There are two types of contributions to the decay amplitude, the first is a Structure Independent (SI), which stems from Inner Bremsstrahlung, that is, the virtual photon is radiated by the $\tau$ and $\pi$ meson where the latter is taken as a point-like particle, therefore, the correspondig matrix element can be obtained using scalar QED. The second is Structure Dependent (SD), which means that the virtual photon can resolve the internal structure of the $\pi$, this means that the matrix element of the weak and the electromagnetic quark currents cannot be factorized into a matrix element of the weak current times a matrix element of the electromagnetic one. This is where one needs to rely on an Effective Field Theory. Separating the left weak current into the vector and axial currents we can express the decay amplitude as follows
 \begin{subequations}\label{eq:amplitudes}
 \begin{align}
     \mathcal{M}_{IB}&=-iG_FV_{ud}fm_\tau\frac{e^2}{k^2}J_\ell^\nu\times\nonumber\\
     &  \overline{u}_{\nu_\tau}(1+\gamma_5)\left[\frac{2p_\nu}{2 p\cdot k+k^2}+\frac{2{p_\tau}_\nu-\slashed{k}\gamma_\nu}{-2p_\tau\cdot k+k^2}\right]u_\tau,\\
     \mathcal{M}_{V}&=-G_FV_{ud}\frac{e^2}{k^2}J_\ell^\nu J_\tau^\mu F_V(W^2,k^2)\varepsilon_{\mu\nu\alpha\beta}k^\alpha p^\beta,\label{eq:FV}\\
     \mathcal{M}_{A}&=iG_FV_{ud}\frac{e^2}{k^2}J_\ell^\nu J_\tau^\mu\times\nonumber\\
     &\left\{F_A(W^2,k^2)\left[(W^2+k^2-m_\pi^2)g_{\mu\nu}-2k_\mu p_\nu\right]\frac{}{}\nonumber\right.\\
     &\left.-\frac{}{}A_2(W^2,k^2)k^2g_{\mu\nu}+A_4(W^2,k^2)k^2(p+k)_\mu p_\nu\right\}\label{eq:FA},
 \end{align}
 \end{subequations}
 where $J_\ell^\nu=\overline{u}(p_-)\gamma^\nu v(p_+)$ and $J_\tau^\mu=\overline{u}(q)(1+\gamma_5)\gamma^\mu u(p_\tau)$ are the lepton electromagnetic and $\tau$ weak charged currents, respectively. The definition of momenta is given in Fig. \ref{fig:SI} Also, $W^2=(p_\tau-q)^2$ and $k = p_+ +p_-$. It is in the Form Factors of eq. (\ref{eq:FV}) and (\ref{eq:FA}) that the hadronic dynamics that cannot be described by means of the underlying theory (QCD) is encoded. These are the expressions that we will compute using R$\chi$T. 
 
 \section{Structure dependent part}
 
 \begin{Figure}
 \centering\includegraphics[width=\linewidth]{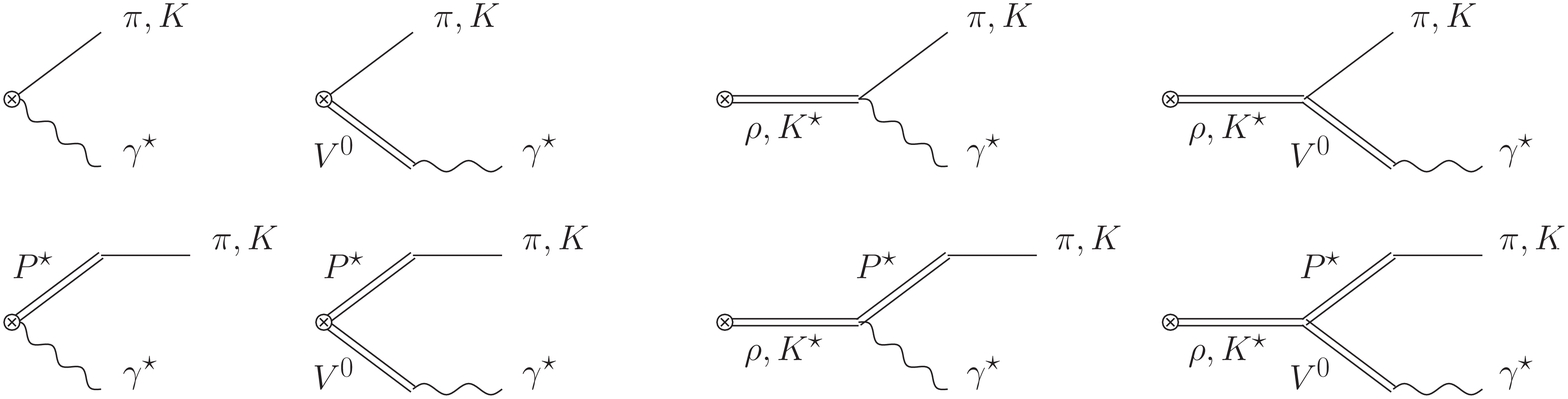}\\
 \footnotesize{Figure 2. Feynman diagrams of the vector SD contributions.}
\end{Figure}

Using R$\chi$T we find that the Feynman diagrams that contribute to the vector form factor are shown in Figure 2, while those contributing to the axial form factors are given in Figure 3. We refer the reader to ref. \cite{Guevara:2021tpy} for the operator basis used. The expression for the vector form factor gives

 \begin{equation}\label{eq:FFV}
     F_V(W^2,k^2)=\frac{1}{3f}\left\{-\frac{N_C\frac{}{}}{8\pi^2}+64 m_\pi^2 C_7^{W\star}\hspace*{14ex}
          \right.\nonumber\end{equation}\begin{equation}
         -8C_{22}^{W}(W^2+k^2)+\frac{4{F_V^{ud}}^2}{M_\rho^2-W^2}\frac{d_3(W^2+k^2)+d_{123}^\star m_\pi^2}{M_\omega^2-k^2}    \nonumber\end{equation}\begin{equation}
\hspace*{-2ex}+\frac{2\sqrt{2}F_V^{ud}}{M_V}\frac{c_{1256}W^2-c_{1235}^\star m_\pi^2-c_{125}k^2}{M_\rho^2-W^2}
          \nonumber\end{equation}\begin{equation}
          \left.\hspace*{4ex}+\frac{2\sqrt{2}F_V^{ud}}{M_V}\frac{c_{1256}k^2-c_{1235}^\star m_\pi^2-c_{125}W^2}{M_\omega^2-k^2}
          \right\},
 \end{equation}

 which is the same result as for the $\pi^0$-TFF (accounting for factors of 2 from Bose symmetry) with the leading-order chiral symmetry breaking terms \cite{Guevara:2018rhj}.

\begin{Figure}
 \centering\includegraphics[width=\linewidth]{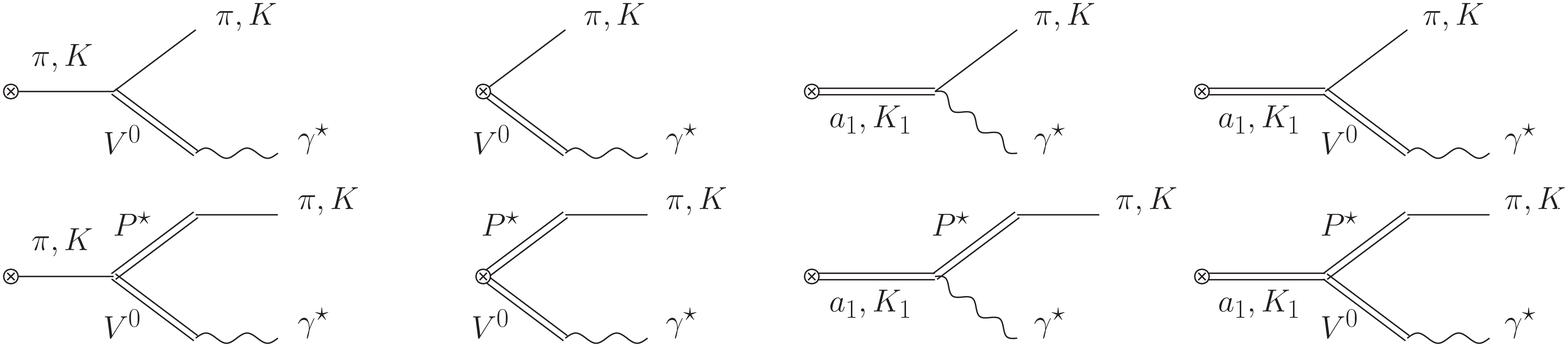}\\
 \footnotesize{Figure 3. Feynman diagrams of the axial SD contributions.}
\end{Figure}

From the diagrams of Figure 3 we find the expressions
\begin{equation}\label{eq:FFA}
     F_A(W^2,k^2)= \frac{F_V^{ud}}{2f}\frac{F_V^{ud}-2G_V-m_\pi^2\frac{4\sqrt{2}d_m}{M_{\pi'}^2}(\lambda_1^{PV}+2\lambda_2^{PV})}{M_\rho^2-k^2}
   \nonumber\end{equation}\begin{equation}
  \hspace*{-5ex}-\frac{F_A}{2f}\frac{F_A-2m_\pi^2\frac{4\sqrt{2}d_m}{M_{\pi'}^2}\lambda_1^{PA}}{M_{a_1}^2-W^2}\nonumber\end{equation}\begin{equation}\hspace*{10ex}                                                                                                         +\frac{\sqrt{2}}{f}\frac{F_AF_V^{ud}}{M_{a_1}^2-W^2}\frac{\lambda_0^\star m_\pi^2-\lambda'k^2-\lambda''W^2}{M_\rho^2-k^2},
 \end{equation}

  \begin{equation}
     A_2(W^2,k^2)=\frac{2}{f}\left(G_V+\frac{2\sqrt{2}m_\pi^2d_m}{M_{\pi'}^2}\lambda_1^{PV}\right.\hspace*{10ex}\nonumber\end{equation}\begin{equation}\hspace*{10ex}\left.+\frac{\sqrt{2}F_A}{M_{a_1}^2-W^2}W^2(\lambda'+\lambda'')\right)\frac{F_V^{ud}}{M_\rho^2-k^2},
 \end{equation}

 \begin{equation}
     A_4(W^2,k^2)=\frac{2}{f}\frac{F_V^{ud}}{M_\rho^2-k^2}\left[\frac{G_V}{W^2-m_\pi^2}\right.\hspace*{12ex}\nonumber\end{equation}\begin{equation}\hspace*{11ex}\left.+\frac{2\sqrt{2}d_mm_\pi^2\lambda_1^{PV}}{M_\pi^2\left(W^2-m_\pi^2\right)}+\frac{\sqrt{2}F_A(\lambda'+\lambda'')}{M_{a_1}^2-W^2}\right].
 \end{equation}

 Taking the chiral limit ($m_\pi\to0$), it can be seen that $A_2$ and $A_4$ recover their linear dependency \cite{Cirigliano:2004ue, Cirigliano:2006hb,Knecht:2001xc}.\\
 
 Imposing the high-energy beahviour of QCD form factors allows us to constrain the parameters of the model. The behavior of the vector form factor in the limits $\lim_{\lambda\to\infty}F_V(\lambda W^2,0)$ and $\lim_{\lambda\to\infty}F_V(\lambda W^2,\lambda k^2)$ 
 \cite{Lepage:1980fj,Brodsky:1973kr} give the constraints (see ref. \cite{Guevara:2018rhj})

 \begin{itemize}
\item $F_V(W^2,k^2)$, $\mathcal{O}(m_P^0)$:
\begin{eqnarray}
C_{22}^W&=&0\,,\label{eq:SD_V_chiral_1}\\
c_{125}&=&0\,,\\
c_{1256}&=&-\frac{N_C M_V}{32\sqrt{2}\pi^2F_V}\,,\\
d_3&=&-\frac{N_CM_V^2}{64\pi^2F_V^2}\,.\label{eq:SD_V_chiral_last}
\end{eqnarray}
\item $F_V(W^2,k^2)$, $\mathcal{O}(m_P^2)$:
\begin{eqnarray}
\lambda_V&=&-\frac{64\pi^2F_V}{N_C}C_7^{W*}\,,\\
c_{1235}^*&=&\frac{N_CM_Ve^V_m}{8\sqrt{2}\pi^2F_V}+\frac{N_CM_V^3\lambda_V}{4\sqrt{2}\pi^2F_V^2}.
\end{eqnarray}
\end{itemize}

We also use the relations fron VVP Green's Function, for the sake of predictability. Also, we use the values for some parameters of the fit done in ref. \cite{Guevara:2018rhj}. These are
$$c_{125}=c_{1235}=0,\hspace*{2ex}c_{1256}=-\frac{N_CM_V}{32\sqrt{2}\pi^2F_V},$$\\
$$\kappa_5^P=0,\hspace*{2ex}d_3=-\frac{N_CM_V^2}{64\pi^2F_V^2}+\frac{F^2}{8F_V^2}+\frac{4\sqrt{2}d_m\kappa_3^{PV}}{F_V},$$\\
\begin{equation}C_7^W=C_{22}^W=0,\hspace*{3ex}d_{123}=\frac{F^2}{8F_V^2}.\end{equation}

However, if we impose that the axial form factors vanish in the same limits as the vector form factors we obtain no constraint since they already have the correct asymptotic behavior. Instead, we use the VAP Green's Function behavior \cite{Cirigliano:2004ue,Cirigliano:2006hb,Knecht:2001xc}, which gives the following relations

\begin{eqnarray}
    \lambda_0=\frac{F^2}{4\sqrt{2}F_VF_A},\hspace*{2ex} d_m\lambda_2^{PV}=\frac{3F^2+2F_A^2-2F_V^2}{16\sqrt{2}F_V},\hspace*{1ex} \nonumber\\
    \lambda''=-\frac{F^2+F_A^2-2F_VG_V}{2\sqrt{2}F_VF_A},\hspace*{2ex}   d_m\lambda_1^{PV}=-\frac{F^2}{4\sqrt{2}F_V},\nonumber\\ \lambda'=\frac{F^2+F_A^2}{2\sqrt{2}F_VF_A}, \hspace*{2ex}d_m\lambda_1^{PA}=\frac{F^2}{16\sqrt{2}F_A}\,.\hspace*{11ex}
\end{eqnarray}

\section{Fit}
Although the spectra of the invariant $m_{\pi e^+e^-}$ mass shows the typical $\rho(1450)-\rho(1700)$ interference (see \cite{Guevara:2021tpy,Jin:2019goo}), 
 involving heavier copies of the $\rho$ meson implies considering more unconstrained parameters, wich cannot be well contrained using the data due to large errors. It is, therefore, expected that some relations stemming from high-energy QCD or Green's Functions will not be fulfilled. Thus, we do not rely on Weinberg's sum rules.\\
 
 Comparing the branching fraction data from Belle with the expected signal events distribution, we obtain an estimation for the deconvolution of signal from the detector, which we do not know. This is taken as a systematic uncertainty, which is comparable to the one reported by Belle. Also, the use of incomplete expressions for the axial form factors by the Belle collaboration to obtain the branching ratios and invariant mass spectrum \cite{Guevara:2021tpy,Jin:2019goo} can lead to biased estimations. We, therefore, choose to fit the total branching fraction to the data instead of computing it from the partial width expression. Thus, we fit $F_V, F_A, \lambda^\star_0$ and $\mathcal{B}$, the branching ratio.\\
 
 The resolution of the data is not precise enough in order to obtain a determined set of parameters that minimize the $\chi^2$ of the fit. Therefore, we fix one of the parameters and fit the other three. We fix first $F_A=130$ MeV, and then $\lambda_0^\star=102\times10^{-3}$. The former is in agreement with the high energy relation of ref. \cite{Roig:2013baa}, while the latter is the estimation of $\lambda_0^\star$ neglecting chiral symmetry breaking effects \cite{Miranda:2020wdg}. A third fit fixing $F_V=\sqrt{3}F$ (its correct high energy value \cite{Roig:2013baa}) was done, however, it furnishes a very poor fit. The fits obtained are shown in Table \ref{tab:fit_par}. As can be seen from set 2 in this Table, $F_V$ is far closer to $\sqrt{2}F$ than to $\sqrt{3}F$, which reiforces the decision of neglecting the Weinberg's sum rules. The spectra obtained from the number of events spectrum for each parameter set along with the experimental data is shown in Figure 4.

 \begin{Figure}
\includegraphics[width=\linewidth]{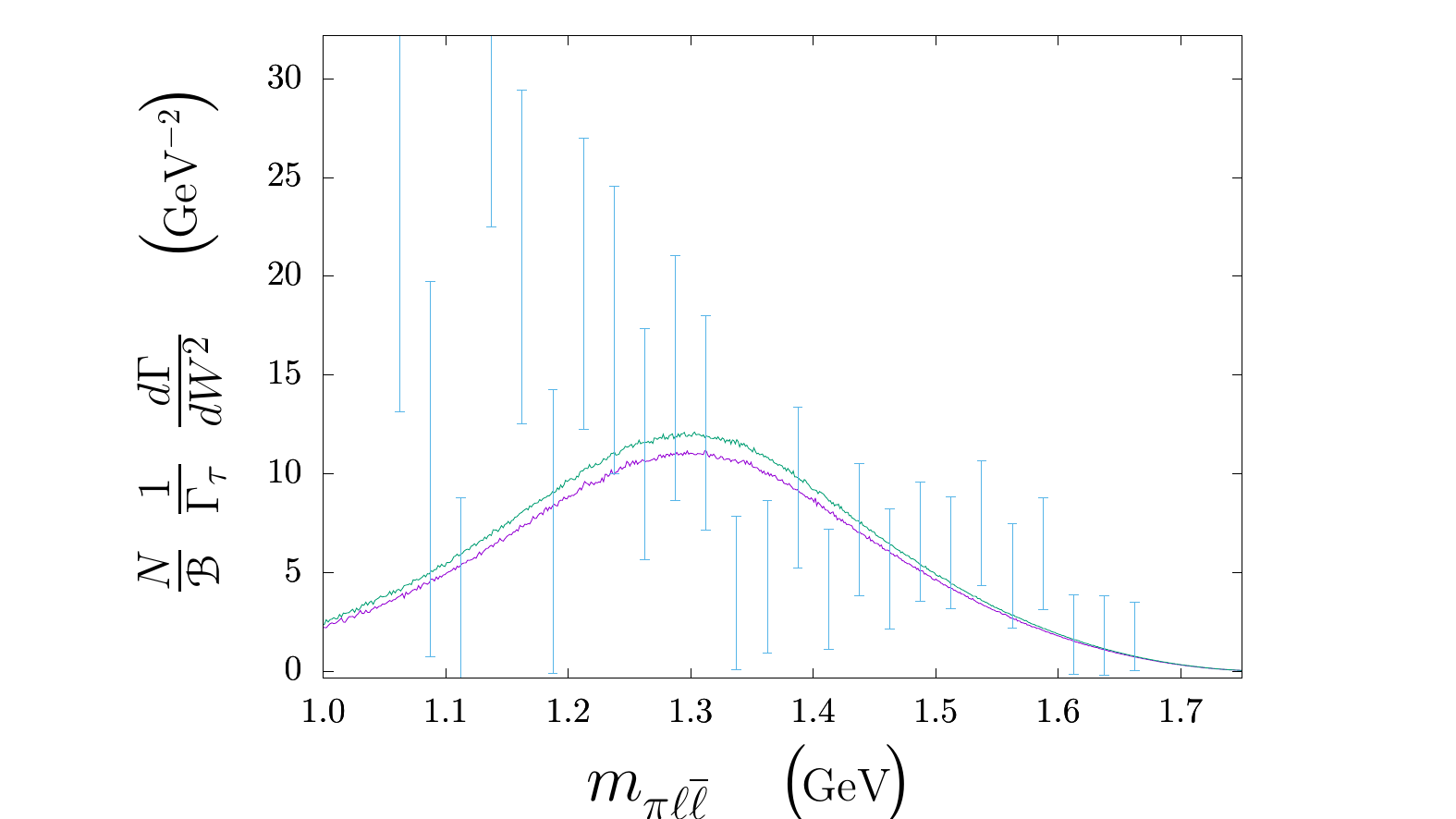}
\footnotesize{Figure 4. Normalized invariant mass spectra obtained with the two sets of parameters obtained from fitting to the Belle data. The purple line corresponds to the data with $F_A$ fixed, while the green one stands for that with $\lambda_0^\star$ fixed. The blue data corresponds to measurements of $\tau^-$ decays, which show best agreement with our model. \cite{Jin:2019goo} }\end{Figure}

\section{Predictions}
 Using the phase space configuration of ref. \cite{Guevara:2021tpy} and the parameters of both sets we obtained the $m_{\pi e^+e^-}$ and $m_{e^+e^-}$ invariant mass spectra and the total branching ratio. 
We computed the $\mathcal{B}(\tau^-\to\nu_\tau\pi^-e^+e^-)$ implementing the kinematical cut used by Belle to obtain their measurement of the branching ratio, $m_{\pi e^+e^-}\geq1.05$ GeV. The difference between our prediction and Belle's measurement is taken as a systematic uncertainty due to unfolding. Thus, taking into account the correlation of the fit, we produced 2400 points using a Gaussian distribution in the parameter space to integrate the branching fraction. We define our prediction of the branching ratio with uncertainty for each decay channel to be the union of the intervals given by the central values with their errors for the prediction thrown by each parameter set, this is, the union of the intervals of Table \ref{tab:BR}; this is given in Table \ref{tab:fit_par}. We also give the $m_{\pi^-e^+e^-}$ invariant mass spectrum in Figure 5 and the $m_{e^+e^-}$ in Figure 6.

\begin{Figure}
\includegraphics[width=\linewidth]{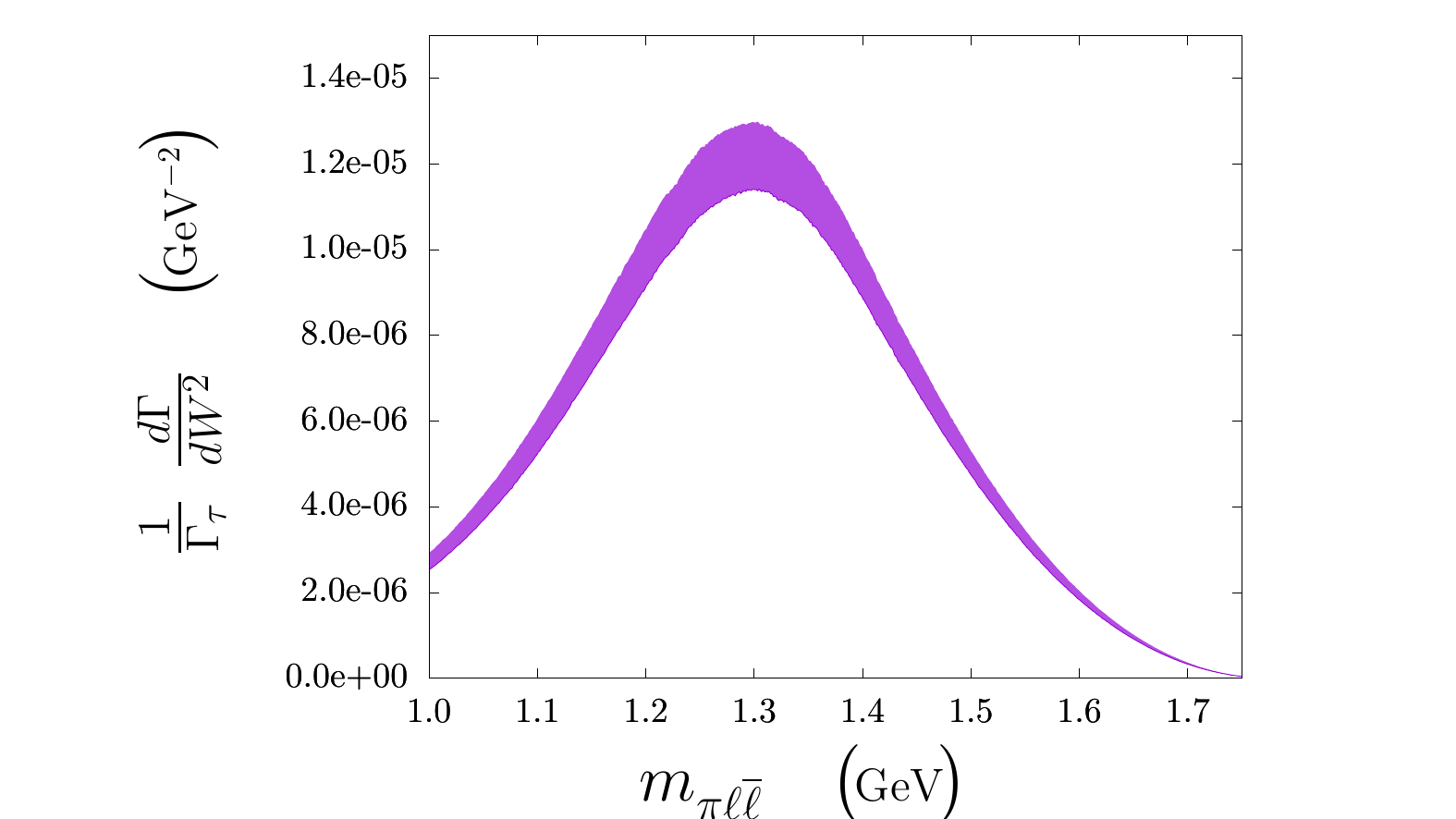}
\footnotesize{Figure 5. Invariant mass spectra $m_{\pi^-e^+e^-}$ for $P=\pi$, the thickness represents the error band obtained from the difference between the two sets.}
\end{Figure}

\begin{Figure}
\includegraphics[width=\linewidth]{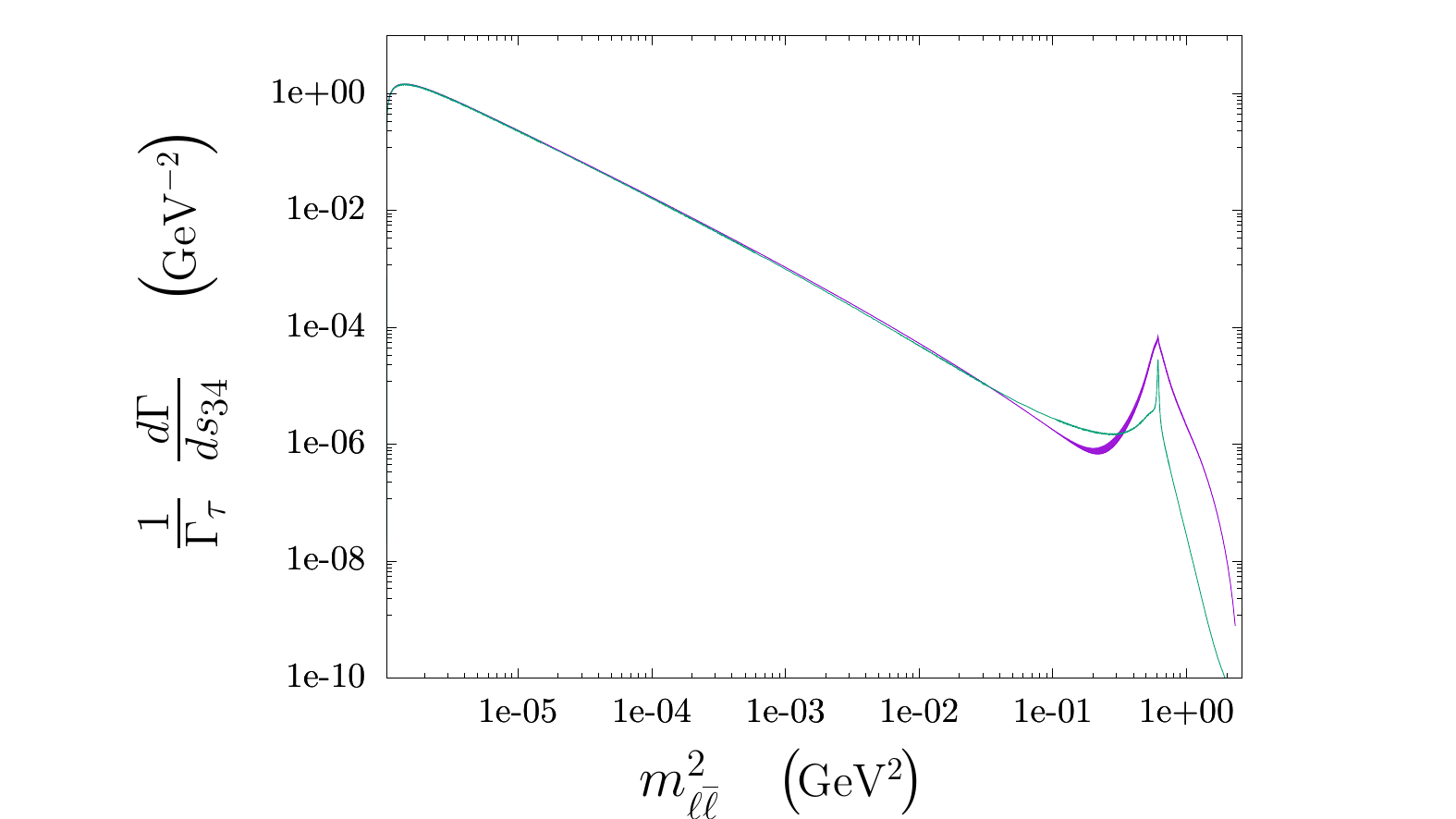}
\footnotesize{Figure 6. Invariant mass spectra $m_{\pi^-e^+e^-}$, the thickness represents the error band obtained from the difference between the two sets. The green line is the prediction of ref. \cite{Guevara:2013wwa}.}
\end{Figure}

\section{Conclusions}
We have given an improved description of the decay amplitude from ref. \cite{Guevara:2013wwa}. This involves a more accurate theoretical description by including flavor breaking terms as well as predictions with reduced uncertainty. As found in the VVP Green's function, the inclusion of pseudoscalar resonances in the form factors is needed to give compatible expressions with the VAP Green's Function analyses \cite{Cirigliano:2004ue,Cirigliano:2006hb,Knecht:2001xc}. A reasonably good fit $\chi^2/dof\approx1.2$ was obtained. However, a better set of data for the $m_{\pi^-e^+e^-}$ and data for the $m_{e^+e^-}$ invariant mass spectra would allow to resolve the dynamics of the heavier copies of the $\rho$ meson involved in these decays. The fact that the fits favor a value $F_V=\sqrt{2}F$ shows that dynamics from such heavier copies are indeed involved, making thus necessary to improve the experimental analyses to obtain more precise data and the need for a $m_{e^+e^-}$ spectra. 

\section{Acknowledgments}

We are indebted to Denis Epifanov and Yifan Jin for leading the Belle analysis of these decays. We thank Pablo S\'anchez Puertas his useful comments on short distance constraints and Juan Jos\'e Sanz Cillero for useful comments on finding the optimal parameter-space basis for the fitting procedure. This work was supported partly by the Spanish MINECO and European FEDER funds (grant FIS2017-85053-C2-1-P) and Junta de Andaluc\'ia (grant FQM-225), partly by the Generalitat Valenciana (grant Prometeo/2017/053) and by C\'atedras Marcos Moshinsky (Fundaci\'on Marcos Moshinsky).

\end{multicols}

\vspace*{1ex}\begin{table}[!ht]
    \centering
    \begin{tabular}{c c c}\hline
    & set 1 & set 2\\
              &  \footnotesize $F_A=130$ MeV & \footnotesize $\lambda_0^\star=102\times10^{-3}$\\\hline\hline
        $F_A$ & 130 MeV& ($122\pm0$) MeV\\
        $F_V$ & (135.5$\pm1.1$) MeV& ($137.4\pm1.6$) MeV\\
        $\lambda_0^\star$ & $(384\pm 0)\times10^{-3}$ & 102$\times10^{-3}$\\
        $\mathcal{B}$ &$(6.01\pm0)\times10^{-6}$ &$(6.36\pm0.12)\times10^{-6}$\\\hline
        $\chi^2/dof$ & 31.1/26& 31.4/26\\\hline\hline
    \end{tabular}
    \caption{Our best fit
    results for $F_A$ , $F_V$, $\lambda_0^*$ and the branching ratio. For the fit  results shown on the left (right) columns we fix $F_A=130$ MeV  ($\lambda_0^\star=102\times10^{-3}$),  respectively. A $0$ error means that the fit uncertainty in the parameter is negligible with respect to its central value.}
    \label{tab:fit_par}
\end{table}

\begin{table}[!ht]
\centering
\begin{tabular}{c c c |c}\hline
      \
     & set 1 & set 2 & IB\\\hline\hline
     $\mathcal{B}(\tau^-\to\nu_\tau \pi^-e^+e^-)$&$(2.38\pm0.28\pm0.11)\cdot10^{-5}$&$(2.45\pm0.45\pm0.04)\cdot10^{-5}$& $1.457(5)\cdot10^{-5}$ \\
      \hline
\end{tabular}\caption{Full branching ratios accounting for both (dominant)  systematic and statistical uncertainties. In the right column we show the SI contribution with the error arising from numerical integration of the differential decay width. }\label{tab:BR}
\end{table}

\begin{table}[!ht]
     \centering
     \begin{tabular}{cc}\hline
          $P,\ell$& $\mathcal{B}(\tau^-\to\nu_\tau P^-\ell\overline{\ell})$ \\ \hline\hline
          $\pi,e$&$ (2.41\pm0.40\pm0.12)\cdot10^{-5}$ \\
          \hline
     \end{tabular}
     \caption{Branching ratios for the different decay channels. The central value is the mean of the union of intervals given in both columns of Table \ref{tab:BR}, the first error covers the width of such union of ranges and the second error is the quadratic mean of statistical uncertainties in Table \ref{tab:BR}.}
     \label{tab:Final_BRs}
 \end{table}
 
 \medline

\begin{multicols}{2}

\end{multicols}
\end{document}